\documentclass[%
 reprint,
 superscriptaddress,
 showpacs,
 aps,
 prl,
]{revtex4-1}

\usepackage[english]{babel}
\usepackage{amsmath,amssymb,amstext,amsfonts}
\usepackage{graphicx}
\usepackage{xfrac}

\graphicspath{{./figs/}}
\DeclareGraphicsExtensions{.pdf,.png}

\renewcommand{\vec}[1]{\boldsymbol{#1}}

\usepackage{hyperref}
\hypersetup{
    colorlinks=true,
    linkcolor=blue,
    filecolor=blue,      
    urlcolor=blue,
    citecolor=blue,
}
 
\usepackage[mathlines,left]{lineno}
\begin{document}


\title{Intrinsic Stabilization of the Drive Beam in Plasma Wakefield Accelerators}

\author{A. Martinez de la Ossa}
\email[]{alberto.martinez.de.la.ossa@desy.de}
\affiliation{Institut f\"ur Experimentalphysik, Universit\"at Hamburg, 22761 Hamburg, Germany}
\author{T.J.~Mehrling}
\email{tjmehrling@lbl.gov}
\affiliation{Deutsches Elektronen-Synchrotron DESY, 22607 Hamburg, Germany}%
\affiliation{Lawrence Berkeley National Laboratory, Berkeley, California 94720, USA}%
\author{J. Osterhoff}
\affiliation{Deutsches Elektronen-Synchrotron DESY, 22607 Hamburg, Germany}%
\email[]{jens.osterhoff@desy.de}

\date{\today}

\begin{abstract}
The hose instability of the drive beam constitutes a major challenge for the stable operation of
plasma wakefield accelerators (PWFAs).
In this work, we show that drive beams with a transverse size comparable
to the plasma blowout radius generate a wake with a varying focusing along the beam,
which leads to a rapid detuning of the slice-betatron oscillations and suppresses the instability.
This intrinsic stabilization principle provides an applicable and effective method
for the suppression of the hosing of the drive beam and allows for a stable acceleration process.
\end{abstract}

\maketitle


In plasma wakefield accelerators (PWFAs),
highly relativistic particle beams are used to excite plasma wakes
which carry extreme accelerating fields~\cite{Veksler1956}.  
The accelerating gradients surpass those produced in today's conventional 
particle accelerators by orders of magnitude and therefore, 
PWFAs constitute an attractive solution for the miniaturization
of the future particle acceleration technology and its derived applications.

Operating PWFAs in the blowout regime~\cite{Rosenzweig1991} enables injection methods
for the production of high-quality witness beams~\cite{Oz2007,Hidding2012,MartinezdelaOssa2013,MartinezdelaOssa2015,Wittig2015,MartinezdelaOssa2017}
and the efficient acceleration within the plasma wake~\cite{Lotov2005,Tzoufras2008}.
However, due to the extreme focusing fields in the blowout plasma cavity,  
the drive and witness beams in PWFAs are subject to transverse instabilities with large growth rates.
In particular, the hose instability (HI) of the drive beam
constitutes a major challenge for the optimal operation of PWFAs~\cite{Huang2007}. 
The HI is initiated by a transverse deviation of the centroid of the drive beam
which causes a displacement of the center of the focusing ion-channel,
which in turn feeds back into the trailing part of the beam,  
leading to the resonant build-up of the transverse centroid oscillations.
It was recently shown that the inherent drive beam energy loss detunes
the betatron oscillations of beam electrons and thereby mitigates the HI~\cite{Mehrling2017}.
Still, for drive beams with a substantial hosing seed,
beam break-up can occur before this mitigation mechanism becomes effective.

In this Letter, we show by means of analytical theory and particle-in-cell (PIC)
simulations with HiPACE~\cite{Mehrling2014},
that drive beams with a transverse size comparable to the plasma blowout radius
generate a wake with a varying focusing along the drive beam,
which causes a rapid detuning of the centroid oscillations and suppresses the HI.
Still, the plasma blowout is completely formed in regions behind the drive beam,
and therefore, the witness beams can be efficiently accelerated with no emittance degradation.
The damping effect caused by head-to-tail variations of the betatron frequency
is well known in radio frequency accelerators~\cite{Balakin1983,Stupakov1997,Bohn2000},  
and it has been recently shown to apply in the linear regime of
plasma wakefield acceleration~\cite{Vieira2014,Lehe2017} for the mitigation of the HI.
In this work, we show for the first time that this stabilization principle
is compatible with the blowout regime for sufficiently wide, high-current and moderate-length drive beams.
The blowout regime is the most common regime in PWFAs, and therefore,
this work is of crucial interest to understand why the hosing of the drive beam
was avoided in FACET~\cite{Hogan2010} and how it can be further suppressed
in future PWFA experiments~\cite{Aschikhin2016,Yakimenko2016,Heinemann2017}.

We start by considering a relativistic electron beam entering
an initially neutral and homogeneous plasma.
As the beam propagates through the plasma,
it expels plasma electrons by means of its space-charge fields,
generating in this way a plasma wakefield which propagates at the velocity of the beam.
The generated wakefields exert a force $\dot{\vec{p}} = - e \vec{W}$
on the beam electrons,
where $\vec{p}$ is the momentum of a beam electron, $e$ the elementary charge,
$\vec{W} = (E_x-cB_y,E_y+cB_x,E_z)$ the wakefield and $c$ the speed of light.
Expressions for the wakefield $\vec{W}$ have been derived in the linear~\cite{Chen1987,Keinigs1987}
and the blowout regime of PWFAs~\cite{Lotov2004,Lu2006}, for axisymmetric drivers and
assuming a quasi-static plasma response. 
The quasi-static approximation assumes that the fields and currents of the beam
are frozen, or quasi-static, during the plasma evolution in the comoving frame, i.e.~$\partial_t \simeq -c\,\partial_\zeta$ for these quantities,
with $\zeta = z - ct$, denoting the comoving variable.
Under this approximation, it is found from Maxwell equations that 
the wakefields satisfy the following relations,
$\partial_x W_z = \partial_\zeta W_x \simeq - (m\omega_p^2/e)\,(j_{p,x}/n_0c)$,
and  $\partial_x W_x \simeq (m\omega_p^2/2e)\,(1 - n_p/n_0 + j_{p,z}/n_0c)$, 
with $\omega_p = \sqrt{n_0 e^2/m\epsilon_{0}}$ the plasma frequency,
$n_0$ and $n_p$ the unperturbed and perturbed plasma electron density, respectively,
and $j_{p,z}$ ($j_{p,x}$) the longitudinal (transverse) plasma electron current.
Ions are assumed to be immobile and the transverse beam current to be negligible.
Beams with an electron density $n_b$ higher than $n_0$ 
expel essentially all plasma electrons
near the propagation axis forming a homogeneous ion cavity,
delimited by a sheath of plasma electrons.
The maximum distance of this sheath with respect to the beam propagation axis
is commonly referred as the blowout radius, $r_\mathrm{bo}$.
Inside this ion cavity (or blowout)
we have that $\partial_x W_z = \partial_\zeta W_x = 0$ and $\partial_x W_x = m \omega_p^2 / 2 e $, 
and the equation of motion for the beam-electrons can be written as
\begin{equation}
\ddot{x} + \frac{\mathcal{E}}{\gamma}\,\dot{x} + \frac{\mathcal{K}}{\gamma}\,x = 0\,,\label{eq:transeqlin}
\end{equation}
where both the focusing strength, $\mathcal{K} \equiv (e/m)\,\partial_x W_x$,
and the rate of energy change, $\mathcal{E} \equiv \dot{\gamma} = - (e/mc)\,W_z$,
are constant for beam electrons at a fixed $\zeta$-position,
and $\gamma \simeq p_z/mc$.
When $n_b < n_0$ the blowout is not complete
and the charge of the ions is partially screened by the plasma electron density,
i.e. $\mathcal{K} \approx \omega_p^2\,(1-(n_p/n_0))/2$,
for a non-relativistic plasma response in the region of the beam.
Assuming $n_p$ constant with the radius for regions sufficiently close to the propagation axis,
Eq.~(\ref{eq:transeqlin}) is still applicable to the beam-electrons
within a partial blowout, where now $\mathcal{K}$ obtains a $\zeta$-dependency through $n_p(\zeta)$. 
Eq.~(\ref{eq:transeqlin}) describes the transverse betatron oscillations
of the beam-electrons, with a frequency $\omega_\beta(t) = \sqrt{\mathcal{K}/\gamma(t)}$.
Given that $\omega_\beta$ is a slowly varying function~\cite{Kostyukov2004},
i.e. $\dot{\omega}_\beta/\omega_\beta^2 = \mathcal{E}/2\sqrt{\mathcal{K}\gamma} \ll 1$,
analytical solutions to Eq.~(\ref{eq:transeqlin}) can be given in the following form 
\begin{equation}
 x(t) = x_0\,A\,\cos{\phi} + \frac{\dot{x}_0}{\omega_{\beta,0}}\,A\,\sin{\phi},\label{eq:transsollin}
\end{equation}
with $\dot{x}_0 = p_{x,0}/m\gamma_0$, the initial transverse velocity of the electron,
$\omega_{\beta,0} = \sqrt{\mathcal{K}/\gamma_0}$, the initial betatron frequency,
$A(t) = (\gamma_0/\gamma(t))^{1/4}$, the amplitude modulation,
and $\phi(t) = \int_0^t \omega_\beta(t')\, \mathrm{d}t'$, the phase advance.
When $\mathcal{K}(\zeta)$ and $\mathcal{E}(\zeta)$ do not change with time,
the phase advance can be written explicitly as 
\begin{equation}
 \phi(t) = 2\frac{\sqrt{\mathcal{K}}}{\mathcal{E}}\,\left( \sqrt{\gamma}-\sqrt{\gamma_0} \right),\label{eq:phase}
\end{equation}
which for $\mathcal{E} \rightarrow 0$ yields $\phi \simeq \omega_{\beta,0} t$.
We now consider an infinitesimal $\zeta$-slice of the drive beam,
with an initial phase-space distribution $f_0(x_0,p_{x,0},\gamma_0)=f_x(x_0,p_{x,0})\,\delta(\gamma_0)$.
Since $\gamma(t) = \gamma_0 + \mathcal{E}t$ for all electrons within the $\zeta$-slice,
it is straightforward to find an equation for the transverse centroid
$X_b(t) \equiv \int x(t) f_x \mathrm{d}x_0 \mathrm{d}p_{x,0}$,
by taking corresponding averages of Eq.~(\ref{eq:transsollin}).
The resulting equation for $X_b$ has the same functional dependence as Eq.~(\ref{eq:transsollin}),
and therefore, the beam centroids also describe betatron oscillations with frequency $\omega_\beta(t)$
and amplitude $\mathcal{A}(t) = A(t)\,\sqrt{X_{b,0}^2+(\dot{X}_{b,0}/\omega_{\beta,0})^2}$,
where $X_{b,0}$ and $\dot{X}_{b,0} \equiv \int \dot{x}_0(t) f_x \mathrm{d}x_0 \mathrm{d}p_{x,0}$ denote
the initial transverse displacement and velocity of the centroid, respectively.

When the drive beam has a small offset in the $x$ direction, $X_b$,
the resulting wakefields develop an asymmetry in the transverse direction.
At first order perturbation, the modified wakefields $W_x'(x)$ can be considered 
identical to the axisymmetric case, but with a certain offset, $X_c$,
with respect to the propagation axis, i.e. $W_x'(x) = W_x(x-X_c)$.
In the blowout regime of PWFA a differential equation for $X_c$ was derived in~\cite{Huang2007},
for a sufficiently narrow drive beam, completely embedded in the ion-cavity:
\begin{equation}
  \partial_\zeta^2X_c + k_c^2\, (X_c-X_b)=0 \,.\label{eq:wcentroid}
\end{equation}
where $k_c = k_p \sqrt{c_\psi(\zeta) c_r(\zeta)/2}$, and $k_p = \omega_p/c$.
The coefficients $c_\psi(\zeta)$ and $c_r(\zeta)$ account for the relativistic motion of electrons
in the blowout sheath and for a $\zeta$-dependence of the blowout radius and the beam current~\cite{Huang2007}.
Eq.~(\ref{eq:wcentroid}) describes the oscillations of $X_c$ driven by
the beam centroid displacements $X_b$.
In turn, the displacement $X_c$ couples back to $X_b$ according to
\begin{equation}
 \ddot{X}_b + \frac{\mathcal{E}}{\gamma}\,\dot{X}_b + \frac{\mathcal{K}}{\gamma}\,(X_b-X_c) = 0\,.\label{eq:xbtranseqlin}
\end{equation}
This set of coupled equations~(\ref{eq:wcentroid}) and (\ref{eq:xbtranseqlin})
has been studied earlier in the ion-channel regime 
(with $k_c = k_p/\sqrt{2}$ and $\mathcal{E} = 0$)~\cite{Whittum1991,Lampe1993},
and for the blowout regime of PWFA~\cite{Huang2007},
assuming perfectly monoenergetic beams and no energy change ($\mathcal{E} = 0$).
These cases are characterized by an exponential growth of $X_b$ and $X_c$
in time and towards the tail of the beam. 
The HI of the drive beam is initiated by a finite centroid displacement of the drive beam $X_{b,0}$,
which is amplified due to a coherent coupling of different $\zeta$-slices of the beam through the plasma.
The effect of a $\zeta$-dependent energy change in the drive beam, $\mathcal{E}(\zeta)$,
has been recently studied in Ref.~\cite{Mehrling2017};
it was shown that hosing saturates as soon as the centroid oscillations
of various $\zeta$-slices become detuned owing to a differing rate of energy change
and/or an initial energy spread.

In this work we extend the study of the HI of the drive beam in PWFAs,
from earlier considerations with narrow beams,
to cases where the initial transverse dimensions of the drive beams 
are comparable to the blowout radius. 
For this analysis we combine PIC simulation results with theoretical considerations,
so as to demonstrate that by controlling the width of the drive beam at the entrance of the plasma,
it is possible to generate a longitudinally varying focusing strength along the drive beam only,
which rapidly detunes the centroid oscillations of different beam slices, thereby suppressing
the HI on a short time scale, on the order of the betatron oscillation period.

For the PIC simulations,
we consider perfectly monoenergetic, highly relativistic drive beams with
an initially tilted Gaussian electron distribution,
which provides a well defined seed to the HI:
$n_b = n_{b,0}\,\exp{[-\zeta^2/2\sigma_z^2]}\,\exp{[(-(x-X_{b,0}(\zeta))^2-y^2)/2\sigma_{x,0}^2]}$.
The beams propagate through a homogeneous plasma with a density such that $k_p \sigma_z = 1$.
At this density, the plasma blowout radius is approximately given by~\cite{Lu2006}
$k_p r_\mathrm{bo} \approx 2~\sqrt{\Lambda_{b,0}}$,
with $\Lambda_{b,0} \equiv 2I_{b,0}/I_A$, 
$I_A = 17.05~\mathrm{kA}$ the Alfv\`en current
and $I_{b,0}$ the peak current of the beam. 
In all the simulations $I_{b,0} = 2.5~\mathrm{kA}$, for which $k_p r_\mathrm{bo} \approx 1.1$.
The transverse (rms) size $\sigma_{x,0}$ is varied from $0.1$ to $0.9~k_p^{-1}$,
and accordingly $n_{b,0}/n_0 = \Lambda_{b,0}/(k_p\sigma_{x,0})^2$
goes from $29$ to $0.36$.
For the narrow cases ($\sigma_{x,0} \ll r_\mathrm{bo}$) the beam is initially
overdense ($n_{b,0}\gg n_0$), while for the wide cases ($\sigma_{x,0} \sim r_\mathrm{bo}$)
it is underdense ($n_{b,0}\lesssim n_0$).
When $\sigma_{x,0} \approx r_\mathrm{bo}$ then $n_{b,0}/n_0 \approx 1/4$.  
See the Supplemental Material~\cite{suppmat} for additional simulation parameters.

\begin{figure}[t]
 \centering
  \includegraphics[width=1.0\columnwidth]{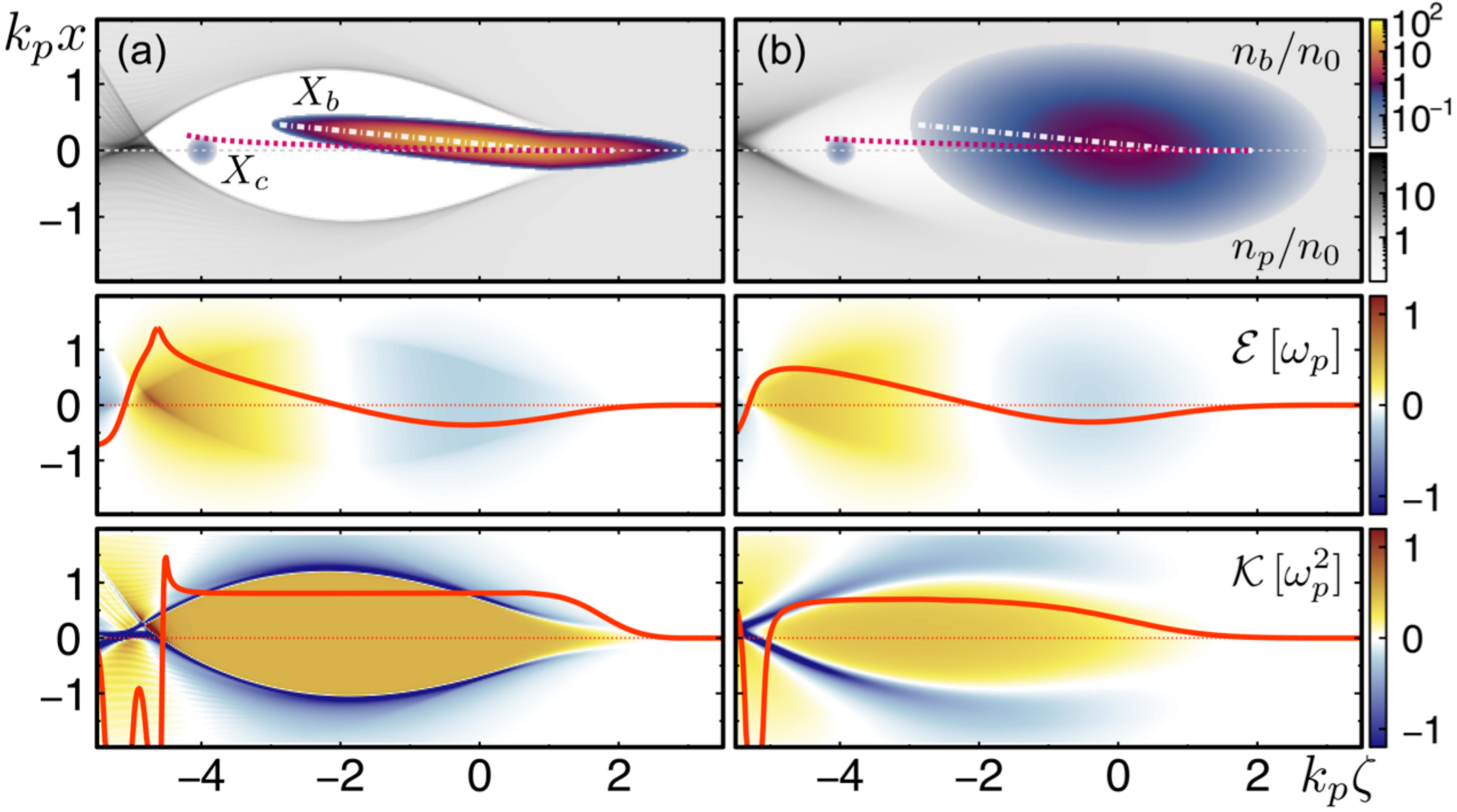}
  \caption{
    PIC simulations for a narrow beam with $k_p\sigma_{x,0} = 0.1$ (a) 
    and a wide beam with $k_p\sigma_{x,0} = 0.5$ (b),
    immediately after entering the homogeneous plasma.
    (Top) Plasma electron density $n_p$ and beam electron density $n_b$. 
    (Middle) Rate of energy change, $\mathcal{E} \equiv -(e/mc)\, E_z$.
    (Bottom) Focusing strength, $\mathcal{K} \equiv (e/m)\, \partial_x W_x$.
    Red curves represent the corresponding lineouts on the propagation axis.
    The centroids of the beam $X_b(\zeta)$ and the focusing channel $X_c(\zeta)$
    are shown in white and purple lines, respectively. 
  }
\label{fig:sim0} 
\end{figure}
Fig.~\ref{fig:sim0} shows the central $\zeta-x$ plane in the beginning of the propagation
in the plasma, for two exemplary simulation runs:
Case $C_a$ with $k_p \sigma_{x,0} = 0.1$ and case $C_b$ with $k_p \sigma_{x,0} = 0.5$.
In case $C_a$, $\sigma_{x,0} \ll r_\mathrm{bo}$ and 
most of the slices of the drive beam are completely embedded in the blowout cavity
(Fig.~\ref{fig:sim0}~(a) - top).
In case $C_b$, the beam is wider and initially underdense, and therefore, 
the blowout formation is only partial in the region of the beam (Fig.~\ref{fig:sim0}~(b) - top).
The energy change along the beam $\mathcal{E}(\zeta)$ is similar for both cases
(Fig.~\ref{fig:sim0}~- middle).
The focusing strength $\mathcal{K}(\zeta)$ along the beam
is perfectly uniform for the narrow beam case $C_a$,
but it substantially varies for the wide beam case $C_b$ (Fig.~\ref{fig:sim0}~- bottom),
where a finite plasma electron density in the region of the beam
alters the focusing field associated with the ion channel. 

\begin{figure}[t]
 \centering
  \includegraphics[width=1.0\columnwidth]{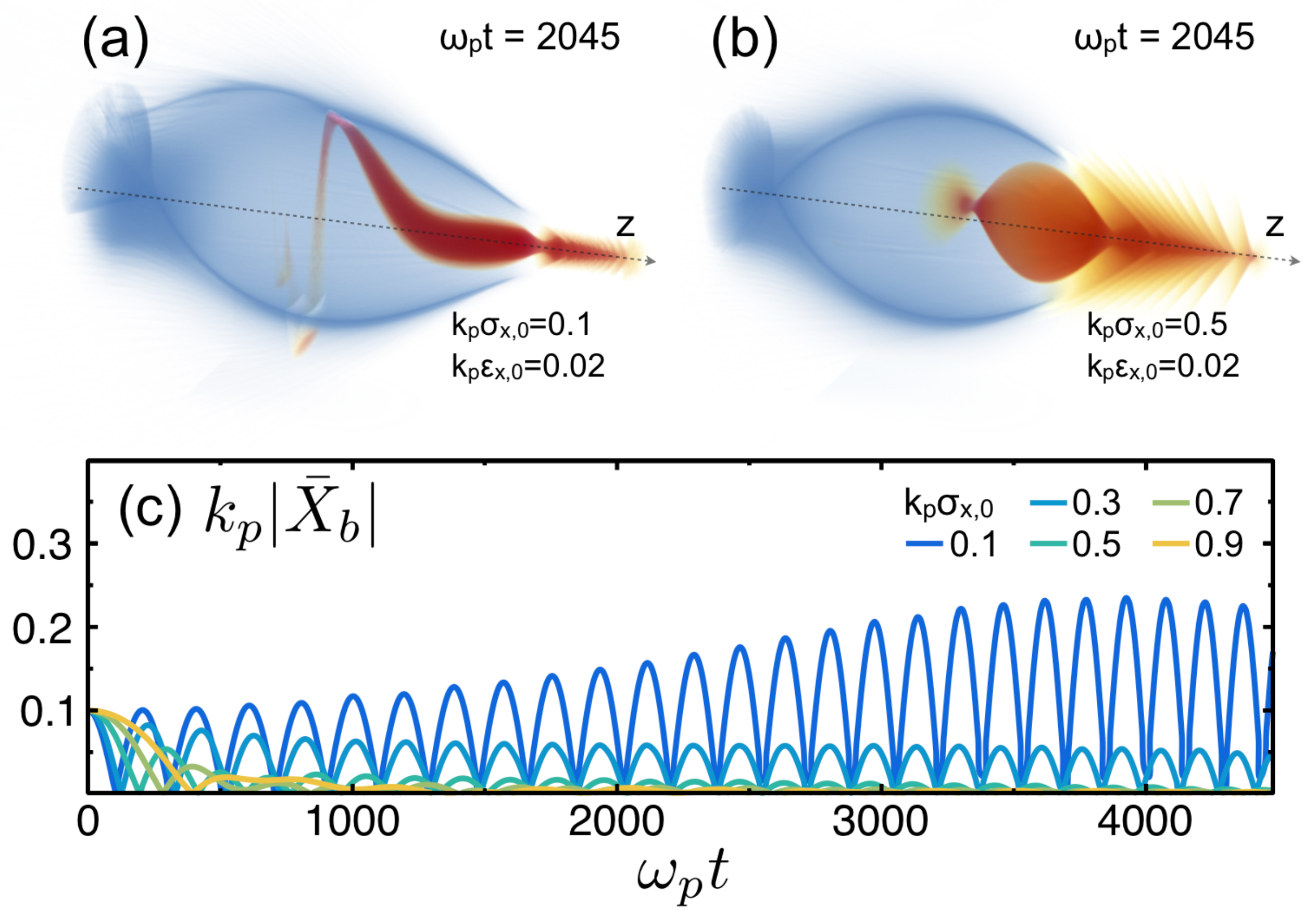}
  \caption{
    PIC simulation results for (a) a narrow beam with $k_p\sigma_{x,0} = 0.1$  
    and (b) a wide beam with $k_p\sigma_{x,0} = 0.5$ (b), after some propagation in the plasma.
    Average centroid oscillations within the central region $k_p\Delta_\zeta=1$ of the drive beam 
    as a function of the propagation time, for five cases with different initial transverse size.
  }
\label{fig:PIChose} 
\end{figure}
The beam and plasma electron densities at $\omega_p t = 2045$
for the cases $C_a$ and $C_b$ are shown in Fig.~\ref{fig:PIChose}(a) and (b), respectively.
After some propagation, the wide drive beam ($C_b$) is transversely compressed
by the self-generated focusing field, enhancing in this way 
the plasma blowout formation (Fig.~\ref{fig:PIChose}(b)). 
The average centroid position $\bar{X}_b$
within a central region of the drive beam with length $k_p\Delta_\zeta=1$,
is shown as a function of the propagation time in Fig.~\ref{fig:PIChose}(c),
for five different initial values of the transverse size (rms). 
It is apparent that the average centroid oscillations are rapidly suppressed
for the cases with a wide beam.
As we explain below, this effect is primarily associated to
a quick decoherence between the oscillations of the slices within the central beam region
due to a non-uniform focusing strength along the drive beam.

We further investigate the stability of the PWFA in the PIC simulations
by studying the evolution of a
low-current witness beam, initially
placed on the propagation axis at comoving position $k_p\zeta = -4$.
The simulations with a narrow drive beam are affected by the HI
and the witness beam breaks up after a short propagation distance.
Only for the wide drive beam cases with $k_p\sigma_{x,0} = 0.7$ and $0.9$,
where the HI is rapidly suppressed, the witness beams are efficiently accelerated
with no slice emittance degradation.
Remarkably, the acceleration performance is barely affected,
dropping only by $10\%$ and $15\%$, respectively, 
when compared to an ideal narrow drive beam case unaffected by hosing.
Extended information about the PIC simulation results can be found on
the Supplemental Material~\cite{suppmat}. 

The decoherence rates owing to longitudinal variations of the betatron frequency
can be estimated by considering an infinitesimal $\zeta$-slice
with constant $\mathcal{K}$ and $\mathcal{E}$, 
together with the solutions of Eq.~(\ref{eq:xbtranseqlin}).
Taking partial derivatives of Eq.~(\ref{eq:phase}),
we obtain the differential phase advance along the beam
\begin{equation}
  \partial_\zeta\phi \simeq  \frac{\omega_{\beta,0} t}{2}\left(\frac{\partial_\zeta\mathcal{K}}{\mathcal{K}}
    -\frac{\partial_\zeta\gamma_0}{\gamma_0}\right)  - \frac{(\omega_{\beta,0}t)^2}{4}\,\frac{\partial_\zeta\mathcal{E}}{\omega_{\beta,0}\gamma_0} ,\label{eq:phasediff}
\end{equation}
where we have included the contribution from a $\zeta$-dependent
initial energy variation in the beam. 
Eq.~(\ref{eq:phasediff}) is valid up to leading order in $t/t_{\mathrm{dp}}$,
with $t_{\mathrm{dp}} \equiv \gamma_0/|\mathcal{E}|$ the energy depletion time.
For an early time, $t \ll t_{\mathrm{dp}}$, 
the phase advance difference between different $\zeta$-slices is dominated by 
either the relative variation of the focusing strength along the beam,
$\kappa \equiv \partial_\zeta\mathcal{K}/\mathcal{K}$,
and/or an initial relative energy chirp, 
which is identically $0$ in the hereby considered cases.
The differential phase advance caused by the variation of $\mathcal{E}$
only appears at second order in $t/t_{\mathrm{dp}}$.

We now consider a beam region with length $\Delta_\zeta$, an uniform current
and with a linear variation of $\mathcal{K}$ and $\mathcal{E}$.
The decoherence time for this beam region can be defined by the time at which
the head-to-tail difference of the phase advance is on the order of $\pi$,
which correspond to opposite oscillation states.
Thus, we use Eq.~(\ref{eq:phasediff}) 
to estimate the decoherence time when either only $\partial_\zeta\mathcal{K} \neq 0$,
i.e. $\omega_{\beta,0}t_{d,\kappa}=2\,\pi/\kappa\Delta_\zeta$, 
or when only $\partial_\zeta\mathcal{E} \neq 0$,
i.e. $\omega_{\beta,0}t_{d,\epsilon} = 2\,\sqrt{\pi/\epsilon \Delta_\zeta}$.
The centroid oscillations of various $\zeta$-slices along the beam region $\Delta_\zeta$
are detuned after the respective decoherence times and the impact of the beam region onto
the focusing channel deviation, which leads to hosing, is strongly suppressed.
As a consequence, the oscillation amplitude of the individual $\zeta$-slices
is expected to saturate and the average centroid displacement within the beam region,
$\bar{X}_b = \Delta_\zeta^{-1}\,\int_{\Delta_\zeta} X_b(\zeta)\,\mathrm{d}\zeta$,
to be strongly damped after the decoherence time.

This model is used to evaluate the decoherence of the centroid oscillations
within a central beam region with length $k_p \Delta_\zeta = 1$ through the quantity $\bar{X}_b$,
for two exemplary cases $C'_a$ and $C'_b$, that resemble the PIC simulation cases
$C_a$, for a narrow beam with $k_p\sigma_{x,0} = 0.1$,
and $C_b$, for a wide beam with $k_p\sigma_{x,0} = 0.5$, respectively.
For simplicity, we assume a fixed channel centroid $X_c = 0$,
and $k_pX_{b,0} = 0.1$, $\dot{X}_{b,0} = 0$ for all the $\zeta$-slices
in the cases $C'_a$ and $C'_b$.
In Fig.~\ref{fig:sim0} we show the values of $\mathcal{E}(\zeta)$ and $\mathcal{K}(\zeta)$
for the PIC simulation cases $C_a$ and $C_b$ in the beginning of the propagation in plasma.
We adopt the central values and derivatives of these quantities in the analytical
calculation of the model cases $C'_a$ and $C'_b$.
In addition, we perform a numerical integration of the exact equation of motion
$\dot{\vec{p}} = - e \vec{W}$, 
for a set of $10^6$ particles representing the considered beam region.
This numerical approach allows to account for non-linear effects
in the motion of the beam electrons with a higher oscillation amplitude, 
which otherwise would not be included in a purely analytical calculation.
The non-uniformity of $\mathcal{K}$ and $\mathcal{E}$ for $|x| \gtrsim r_\mathrm{bo}$
is also accounted for by adopting the values from the PIC simulations (cf. Fig.~\ref{fig:sim0}).

\begin{figure}[t]
 \centering
  \includegraphics[width=1.0\columnwidth]{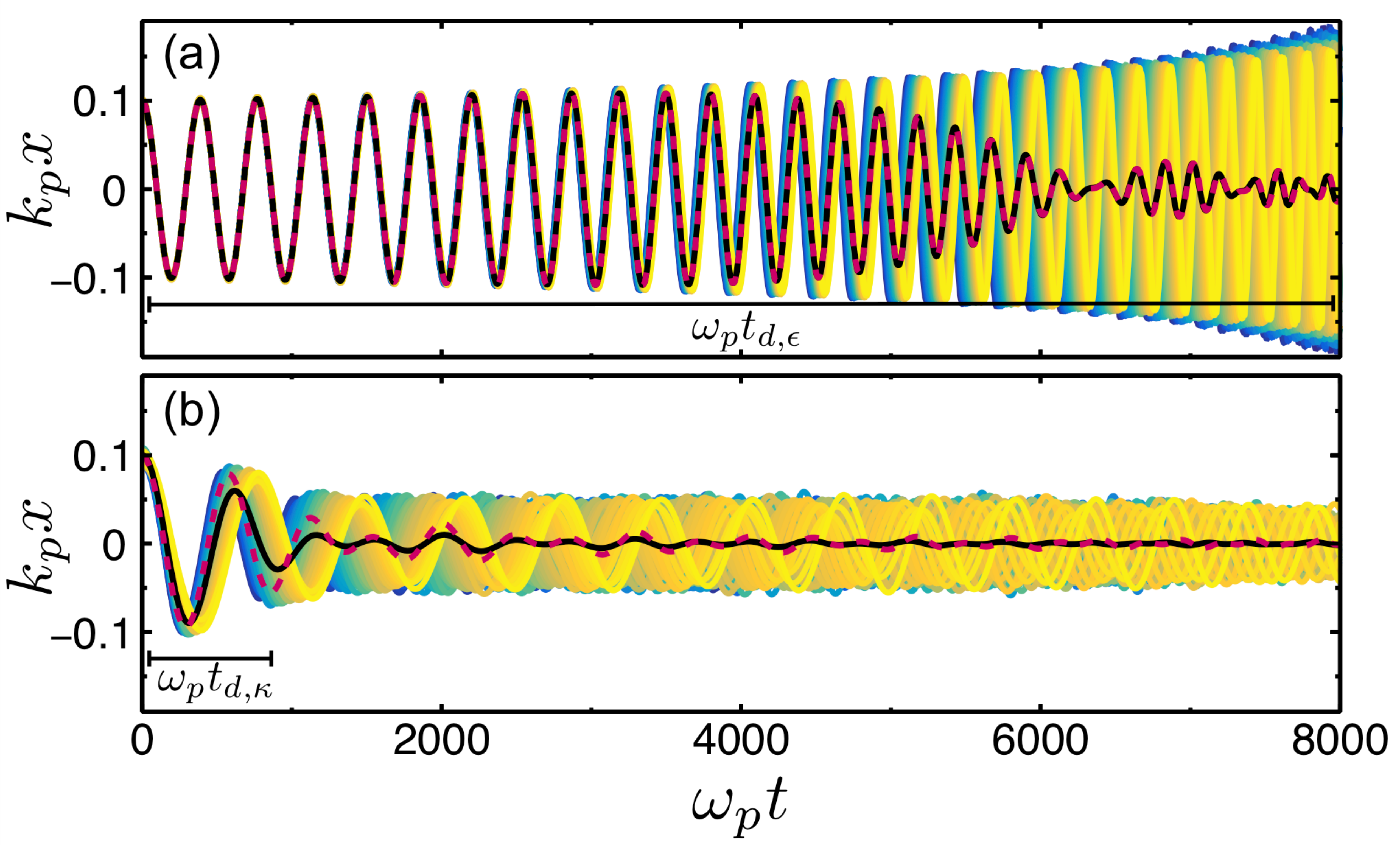}
  \caption{
    Centroid displacements of 50 equally spaced $\zeta$-slices within the beam region
    $k_p\Delta_\zeta=1$ for a narrow beam with $k_p\sigma_{x,0} = 0.1$ (case $C'_a$) (a)
    and a wide beam with $k_p\sigma_{x,0} = 0.5$ (case $C'_b$) (b).
    The centroids are calculated by numerical integration of the
    equations of motion for a set of $10^6$ particles composing the beam region.
    Yellow curves refer to slices near the front and blue curves slices at the back of the beam
    region.
    The black curve shows the average centroid displacement of the beam region, $\bar{X}_b$.
    The red dashed curve represents the analytical calculation for $\bar{X}_b$,
    when just Eq.~(\ref{eq:xbtranseqlin}) with $X_c = 0$
    for the beam centroid displacements is considered. 
  }
\label{fig:deco} 
\end{figure}
In Fig.~\ref{fig:deco} we show the centroid oscillations for 50 $\zeta$-slices
along the considered beam region $\Delta_\zeta$ (colored curves),
together with their average $\bar{X}_b$ obtained from the numerical approach (black line)
and as a result of the analytical model (red dashed line).
For case $C'_a$ (Fig.~\ref{fig:deco} (a)),
$\kappa \simeq 0$ within the considered beam region and the decoherence
occurs predominantly from a differential energy change along the beam.
In this case, the decoherence time is approximately $t_{d,\epsilon} \simeq 8000/\omega_p$, 
which is comparable to the energy depletion time $t_\mathrm{dp} \simeq 9000/\omega_p $.
The analytical model is in excellent agreement with the numerical calculation
for this narrow beam scenario.
For case $C'_b$ (Fig.~\ref{fig:deco} (b)),
$\kappa \neq 0$ and the decoherence from a variation of
the focusing strength along the beam region dominates.
Hence, the decoherence time can be estimated by $t_{d,\kappa} \simeq 800/\omega_p$,
which is on the order of the initial betatron period of the beam electrons
$T_{\beta,0} = 2\pi/\omega_{\beta,0} \simeq 590/\omega_p$. 
In this case, the model predicts that decoherence is reached on
a much shorter time scale than for the narrow beam case $C'_a$,
in good qualitative agreement with the behavior observed in
the PIC simulation cases $C_a$ and $C_b$.

We note that for the wide beam case $C'_b$,
the non-linear effects on the motion of the electrons with a higher oscillation
amplitude cause additional decoherence through intra-slice phase mixing,
and consequently, a damping of the centroid oscillation amplitude of the different $\zeta$-slices.
As a result, the numerical calculation predicts a slightly higher damping of
$\bar{X}_b$ than the analytical model in case $C'_b$ (Fig.~\ref{fig:deco}~(b)).
From the comparison between the analytical and the numerical approaches, 
we identify the decoherence caused by a finite $\partial_\zeta\mathcal{K}$ 
as the main effect responsible for the fast suppression of the HI observed in 
PIC simulations with wide drive beams.

In conclusion, we show that the HI in PWFAs is rapidly suppressed
for drive beams with an initial transverse size comparable to the blowout radius.
The intrinsic variation of the focusing strength in the beam region 
for scenarios with initially wide and underdense drive beams 
leads to a quick decoherence between the centroid oscillations of various
slices along the beam, and consequently, to the suppression of the instability.
Still, behind the drive beam the blowout formation is complete and
the witness beams are efficiently accelerated with no emittance degradation.
This intrinsic stabilization principle provides an applicable and effective method for
the suppression of the HI of the drive beam and will allow for a stable acceleration
process in future PWFA experiments.

\acknowledgements
We acknowledge the grant of computing time by the
J\"{u}lich Supercomputing Center on JUQUEEN under Project No.~HHH23
and the use of the High-Performance Cluster (Maxwell) at DESY.
T.J.M~acknowledges the support by the DAAD with funds from the 
BMBF and the MSCA of the EU's FP7 under REA grant no.~605728 (P.R.I.M.E.) 
and the support by the Director, Office of Science, Office of High Energy Physics, 
of the U.S.~Department of Energy under Contract No.~DE-AC02-05CH11231.
A.M.~acknowledges V. Libov and J. Zemella for useful discussions
in the context of start-to-end simulations for FLASHForward, 
and the Helmholtz Virtual Institute VH-VI-503, for financial support.


%



\pagebreak

\onecolumngrid
\begin{center}
  \textbf{\large Supplemental material: Simulation Setup and Extended Results}\\[.2cm]
  A. Martinez de la Ossa,$^{1,*}$ T.J.~Mehrling,$^{2,3}$ and J. Osterhoff$^2$\\[.1cm]
  {\itshape ${}^1$Institut f\"ur Experimentalphysik, Universit\"at Hamburg, 22761 Hamburg, Germany\\
  ${}^2$Deutsches Elektronen-Synchrotron DESY, 22607 Hamburg, Germany\\
  ${}^3$Lawrence Berkeley National Laboratory, Berkeley, California 94720, USA\\}
  ${}^*$alberto.martinez.de.la.ossa@desy.de\\
(Dated: \today)\\[1cm]
\end{center}
\twocolumngrid

\setcounter{equation}{0}
\setcounter{figure}{0}
\setcounter{table}{0}
\setcounter{page}{1}
\renewcommand{\theequation}{S\arabic{equation}}
\renewcommand{\thefigure}{S\arabic{figure}}
\renewcommand{\bibnumfmt}[1]{[S#1]}
\renewcommand{\citenumfont}[1]{S#1}

\subsection{Simulation Setup}
For the PIC simulations, the quasi-static code HiPACE~\cite{Mehrling2014-S} is used 
together with the following physical setup.
We consider perfectly monoenergetic, highly relativistic drive beams with
$\gamma_0 = 1~\mathrm{GeV}/mc^2$ and with
the following electron density profile:
\begin{equation}
  n_b = n_{b,0}\,\exp{\left[-\frac{\zeta^2}{2\sigma_z^2}-\frac{(x-X_{b,0}(\zeta))^2+y^2}{2\sigma_{x,0}^2}\right]}\,,
  \label{eq:density}
\end{equation}
where $n_{b,0}/n_0 = (2I_{b,0}/I_A)\,/(k_p\sigma_{x,0})^{2}$,
with $I_{b,0} = 2.5~\mathrm{kA}$, the peak current of the beam
and $I_A = 17.05~\mathrm{kA}$, the Alfv\`en current.
The beam-centroid is initially linearly tilted in the $x$ direction, starting at $\zeta = \sigma_z$,
such that $X_{b,0}(\zeta) = - 0.1\,(\zeta-\sigma_z)$ for $\zeta < \sigma_z$ and $X_{b,0} = 0$ otherwise.
The spatial tilt in the $x$ direction provides a well defined seed to the hosing instability.
The initial transverse phase-space distribution of the beams is Gaussian in both $x$ and $p_x$.
The initial transverse (rms) size of the beam, $\sigma_{x,0}$, is varied in the simulations
from $0.1~k_p^{-1}$ to $0.9~k_p^{-1}$, in combination with its initial emittance,
$\epsilon_{x,0} = \sqrt{\langle x_0^2 \rangle \langle p_{x,0}^2 \rangle - \langle x_0 p_{x,0} \rangle^2}/mc$, which ranges from $0.02~k_p^{-1}$ to $0.32~k_p^{-1}$.
Initially, the beams are at waist, i.e $\langle x_0 \, p_{x,0} \rangle = 0$.
In addition, a short, $k_p\sigma_{z,0} = 0.1$, narrow, $k_p \sigma_{x,0} = 0.1$
and perfectly mono-chromatic low-current witness beam is added to the simulations,
in order to test the acceleration performance.
The considered witness beam has the same initial energy and emittance as the drive beam
and it is initially placed on the propagation axis at comoving position $k_p\zeta = -4$.
The simulations use a moving window propagating at $c$.
The dimensions of the simulation box are $9\times7\times7~k_p^{-3}$.
The simulation box is divided into $512\times256\times256$ cells, 
which gives cell sizes of $k_p \Delta \zeta = 0.0176$ and
$k_p \Delta x = k_p \Delta y = 0.0273$.
Each cell contains $2\times2\times2$ simulation particles for the beam electrons
and $1\times2\times2$ particles for the plasma electrons.
The time step for the calculation of the electromagnetic fields is $\Delta t = 5~\omega_p^{-1}$,
which is much smaller that the inverse of the maximum betatron frequency of the
beam electrons $\omega_{\beta,\mathrm{max}}^{-1} = \sqrt{2\gamma_0}\,\omega_p^{-1} \simeq 62\,\omega_p^{-1}$.

\subsection{Extended Results}
\begin{figure*}[h]
 \centering
  \includegraphics[width=2.0\columnwidth]{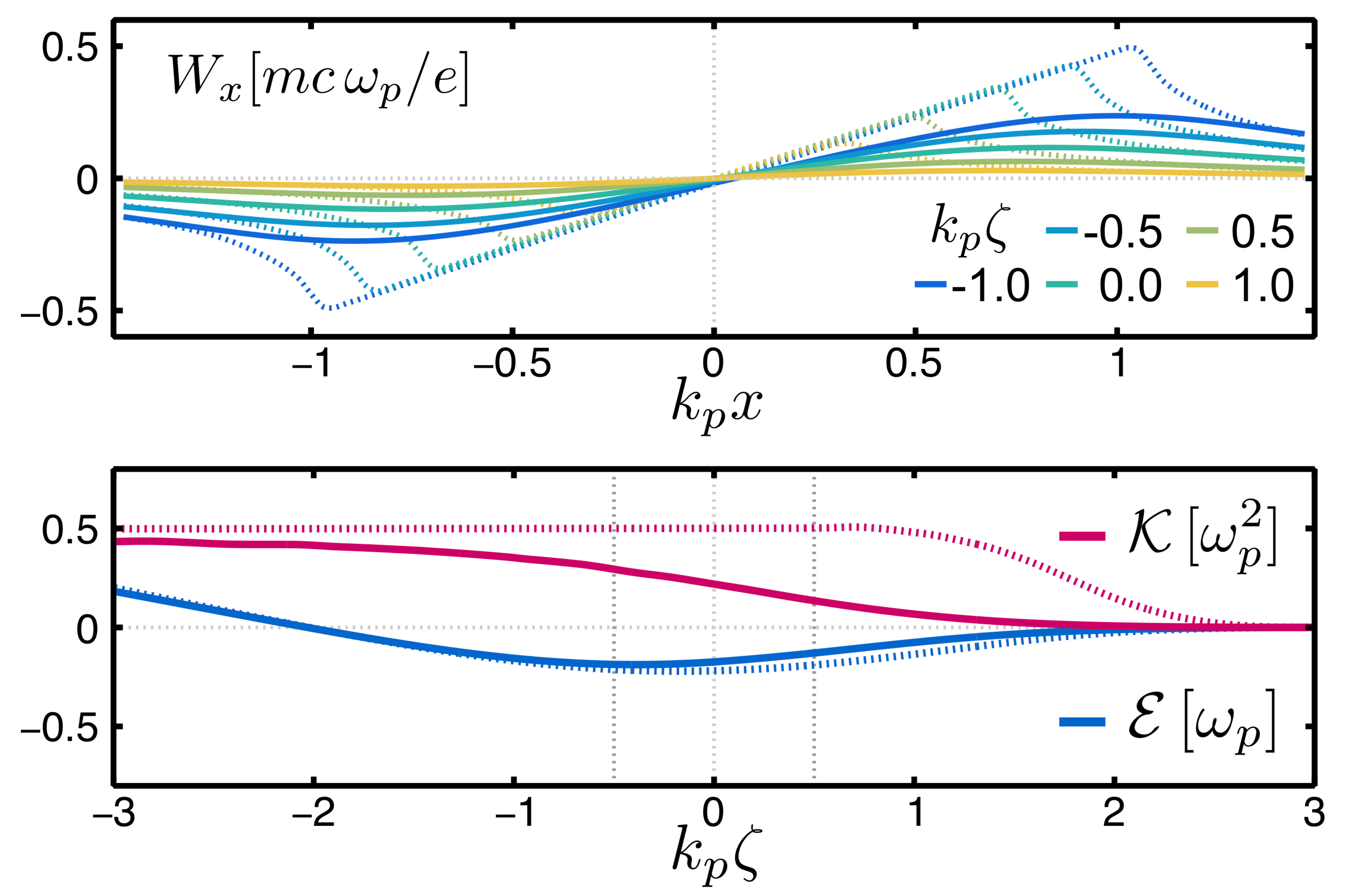}
  \caption{
    (Top) Transverse wakefield ($W_x = E_x - c B_y$) for five different
    $\zeta$-slices along the drive beam.
    (Bottom) Focusing strength $\mathcal{K}$ (purple line) and rate of energy change $\mathcal{E}$
    (blue line) along the drive beam in the beginning of the propagation in plasma,
    for a narrow beam with $k_p\sigma_{x,0} = 0.1$ (dashed lines) and a wide beam with
    $k_p\sigma_{x,0} = 0.5$ (solid lines).
  }
\label{fig:FIG1} 
\end{figure*}

In this section we present a series of figures showing extended information from PIC simulations.
In addition to Fig.~\ref{fig:sim0},
which shows the central $\zeta-x$ plane in the beginning of the propagation
in the plasma, for two exemplary simulation runs:
Case $C_a$ with $k_p \sigma_{x,0} = 0.1$ and case $C_b$ with $k_p \sigma_{x,0} = 0.5$,
both with $k_p\epsilon_{x,0} = 0.02$, 
we have added here Fig.~\ref{fig:FIG1}, which shows the transverse lineouts
of the focusing field $W_x$ for five comoving positions along the drive beam (top panel),
and the longitudinal lineouts of $\mathcal{E}$ and $\mathcal{K}$ along the propagation axis
(bottom panel).
The energy change along the beam $\mathcal{E}(\zeta)$ is similar for both cases.
The focusing strength $\mathcal{K}(\zeta)$ along the beam is perfectly uniform for the narrow beam case $C_a$,
but it substantially varies for the wide beam case $C_b$.

\begin{figure*}[h]
 \centering
  \includegraphics[width=2.0\columnwidth]{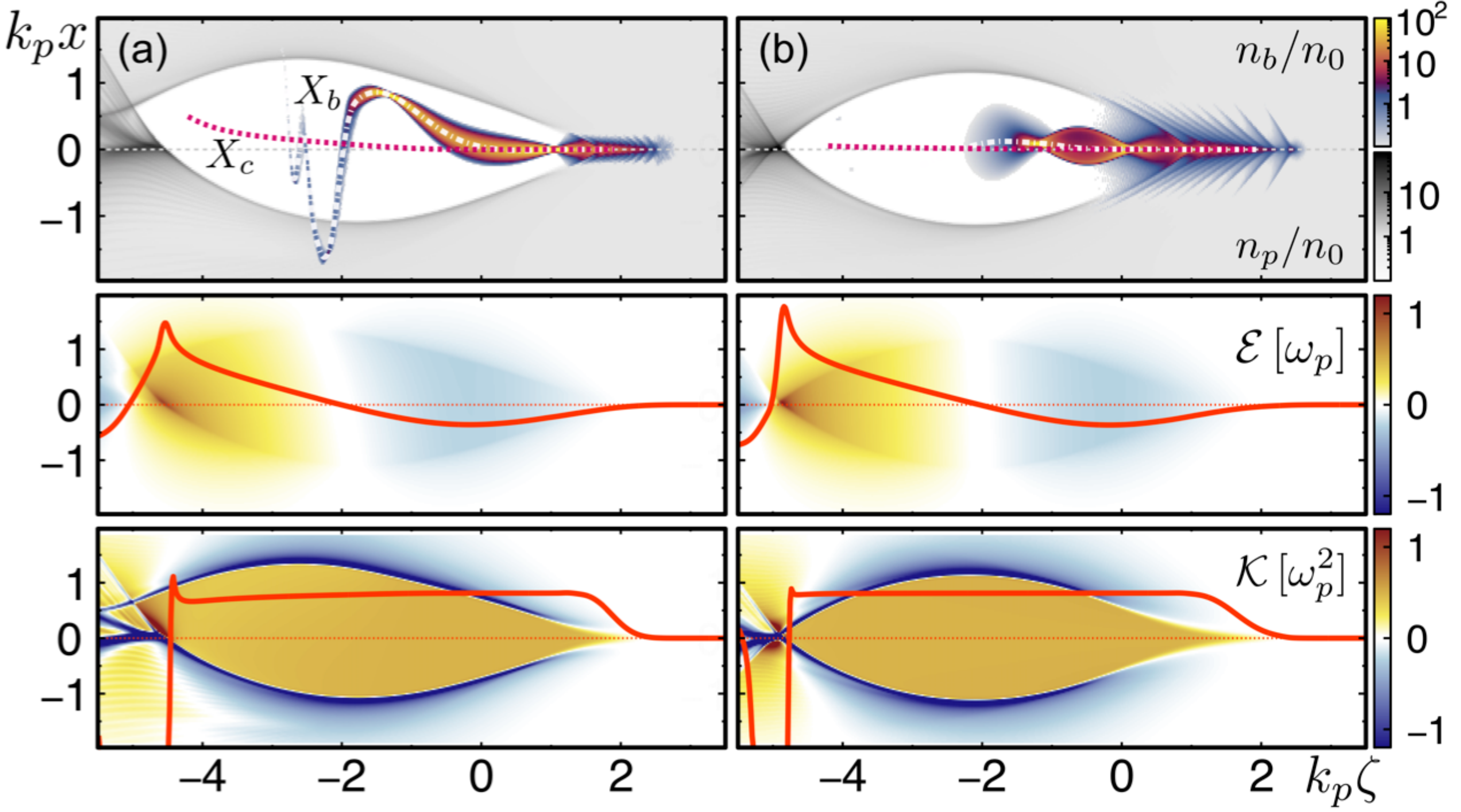}
  \caption{
    PIC simulations for a narrow beam with $k_p\sigma_{x,0} = 0.1$
    (a) and a wide beam with $k_p\sigma_{x,0} = 0.5$ (b), after a propagation time $\omega_p t = 2000$.
    (Top) Plasma electron density $n_p$ and beam electron density $n_b$.
    (Middle) Rate of energy change, $\mathcal{E} \equiv -(e/mc)\, E_z$.
    (Bottom) Focusing strength, $\mathcal{K} \equiv (e/m)\, \partial_x W_x$.
    Red curves represent the corresponding lineouts on the propagation axis.
    The centroids of the beam $X_b(\zeta)$ and the focusing channel $X_c(\zeta)$ are shown
    in white and purple lines, respectively.
  }
\label{fig:FIG2}
\end{figure*}

\begin{figure*}[t]
 \centering
  \includegraphics[width=1.8\columnwidth]{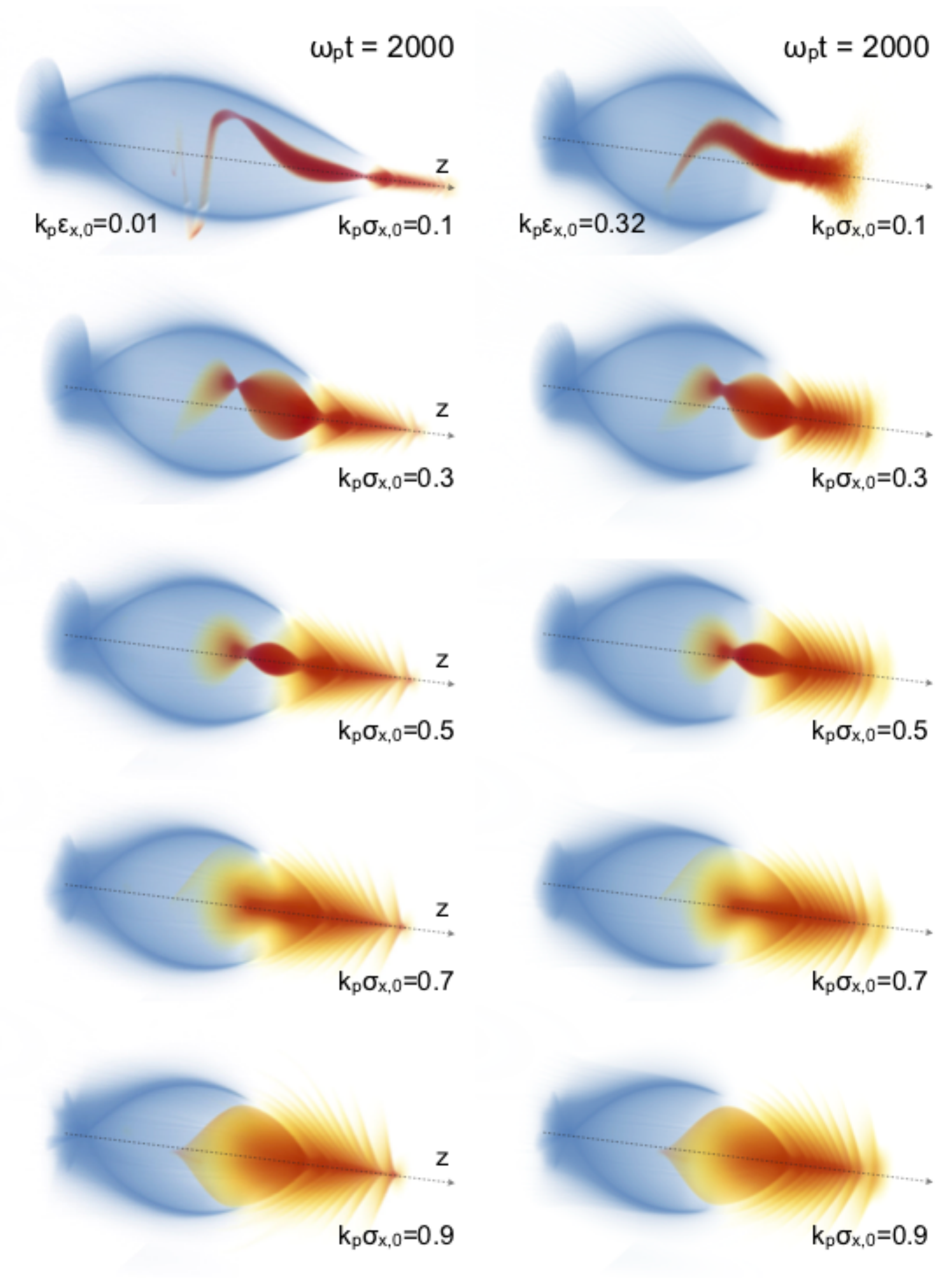}
  \caption{
    PIC simulation results after a propagation time $\omega_p t = 2000$.
    Beam (red) and plasma (blue) electron density for five simulation cases
    with $k_p \epsilon_{x,0} = 0.02$ (left column), and for five simulation cases
    with $k_p \epsilon_{x,0} = 0.32$ (right column), 
    ranging from $k_p\sigma_{x,0} = 0.1$ to $k_p\sigma_{x,0} = 0.9$ (from top to bottom).
  }
\label{fig:FIG3} 
\end{figure*}

Fig.~\ref{fig:FIG2} shows the central $\zeta-x$ plane
after a propagation time $\omega_pt = 2000$, for the same exemplary simulation runs $C_a$ and $C_b$.
It is apparent that, due to the self-generated focusing field,
the wide drive beam is transversely focused, and therefore, the blowout formation is enhanced.
Fig.~\ref{fig:FIG3} shows the beam and plasma densities after a propagation time $\omega_pt = 2000$
for five simulation cases with $k_p \epsilon_{x,0} = 0.02$ (left column) and five simulation cases
with $k_p \epsilon_{x,0} = 0.32$ (right column).
The initial transverse size (rms) of the drive beam is increased from top to bottom,
ranging from $k_p\sigma_{x,0} = 0.1$ to $k_p\sigma_{x,0} = 0.9$ in steps of $0.2$.
Here we see again that the blowout formation behind the driver is complete,
also for the initially wide drive beams, barely differing from the narrow cases.

\begin{figure*}[t]
 \centering
  \includegraphics[width=1.8\columnwidth]{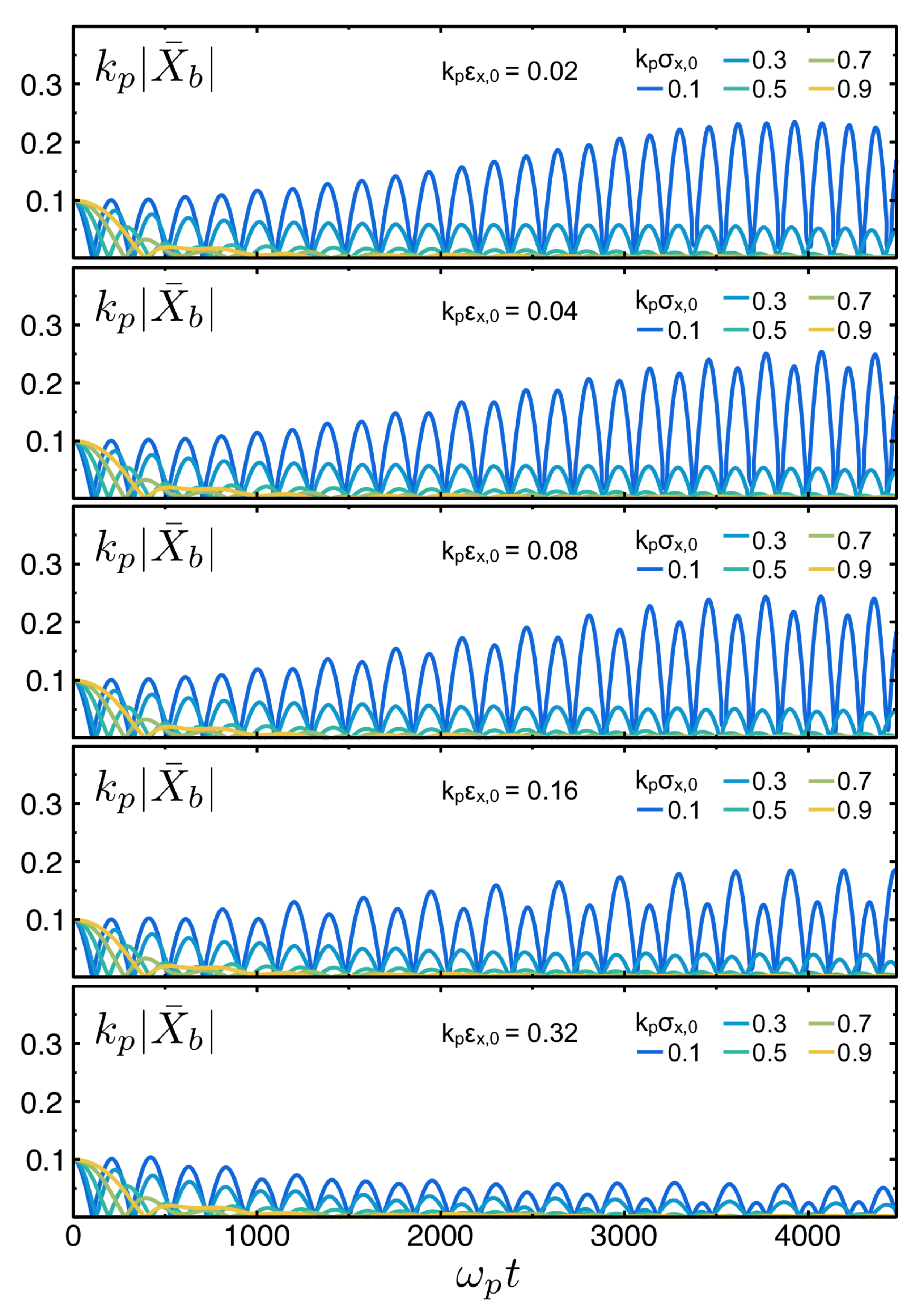}
  \caption{
    Average centroid oscillations within the central region $k_p\Delta_\zeta=1$ of the drive beam 
    as a function of the propagation time, for five PIC simulation cases with different
    initial transverse size and same initial emittance, which ranges from $k_p \epsilon_{x,0} = 0.02$
    to $k_p \epsilon_{x,0} = 0.32$ (from top to bottom).
  }
\label{fig:FIG4} 
\end{figure*}

Fig.~\ref{fig:FIG4} shows the average centroid oscillations of the central region
of the drive beam when its initial emittance is increased
with respect to the narrow case considered in the article, i.e. with $k_p \epsilon_{x,0} = 0.02$.
For the wide drive beam cases with $k_p\sigma_{x,0} \gtrsim 0.5$, the average transverse
oscillations of the beam are strongly damped after a short propagation time,
for any of the considered values of the initial emittance.
The effect of a higher initial emittance is only relevant for the narrow drive beam cases.
We observe that, for the narrow beam case with $k_p\sigma_{x,0}=0.1$ and $k_p \epsilon_{x,0} = 0.32$,
the head of the beam expands transversely (cf. Fig.~\ref{fig:FIG3}~upper-right corner) and,
after few betatron oscillations, the central beam region is affected by
the decoherence associated to a non-uniform $\mathcal{K}$,
causing the damping of the average centroid oscillations.

\begin{figure*}[t]
 \centering
  \includegraphics[width=1.8\columnwidth]{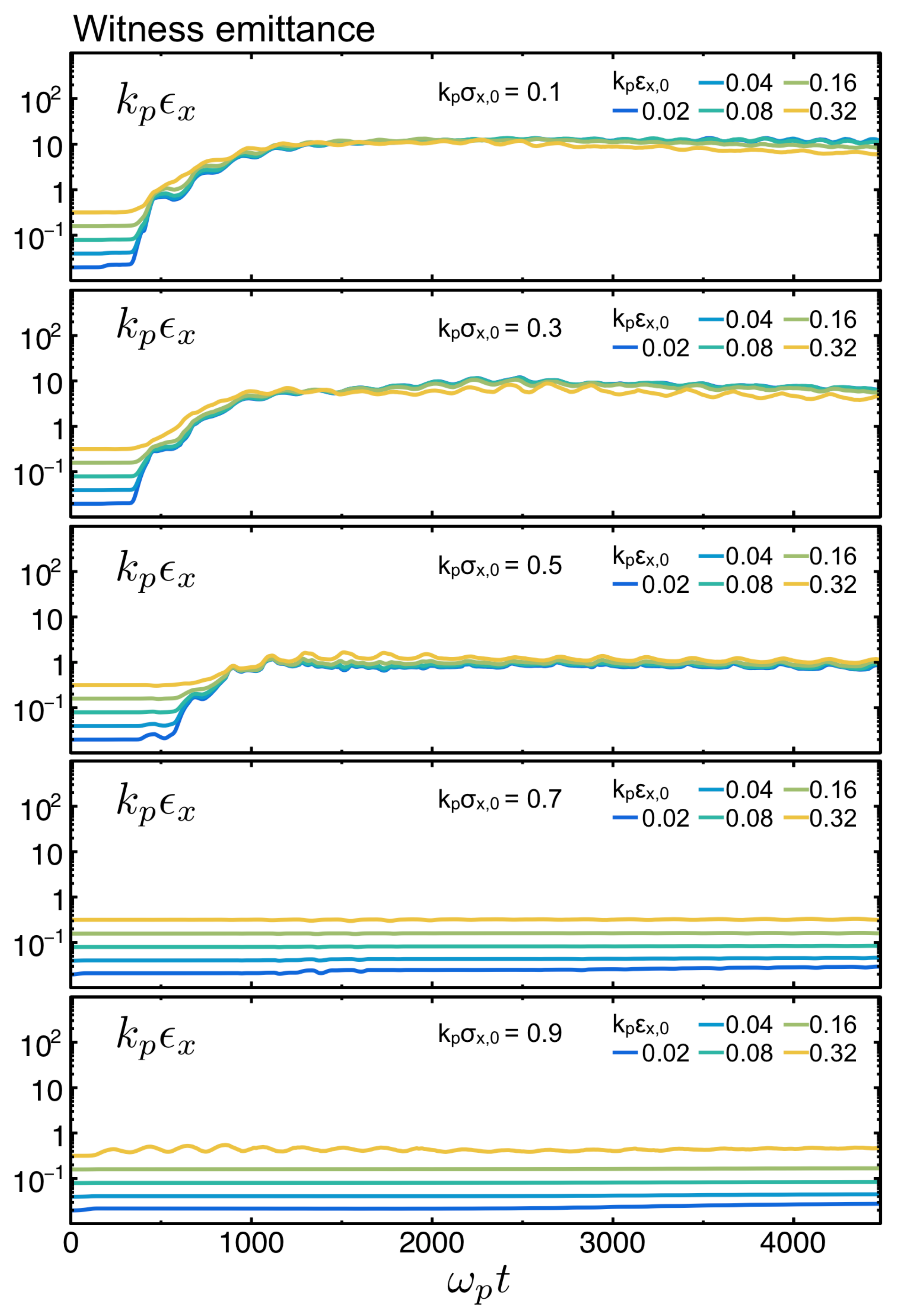}
  \caption{
    Average slice emittance of the witness beam as a function of the propagation time,
    for five PIC simulation cases with different initial emittance and same initial transverse size,
    which ranges from $k_p\sigma_{x,0} = 0.1$ to $k_p\sigma_{x,0} = 0.9$ (from top to bottom).
  }
\label{fig:FIG5} 
\end{figure*}

\begin{figure*}[t]
 \centering
  \includegraphics[width=1.8\columnwidth]{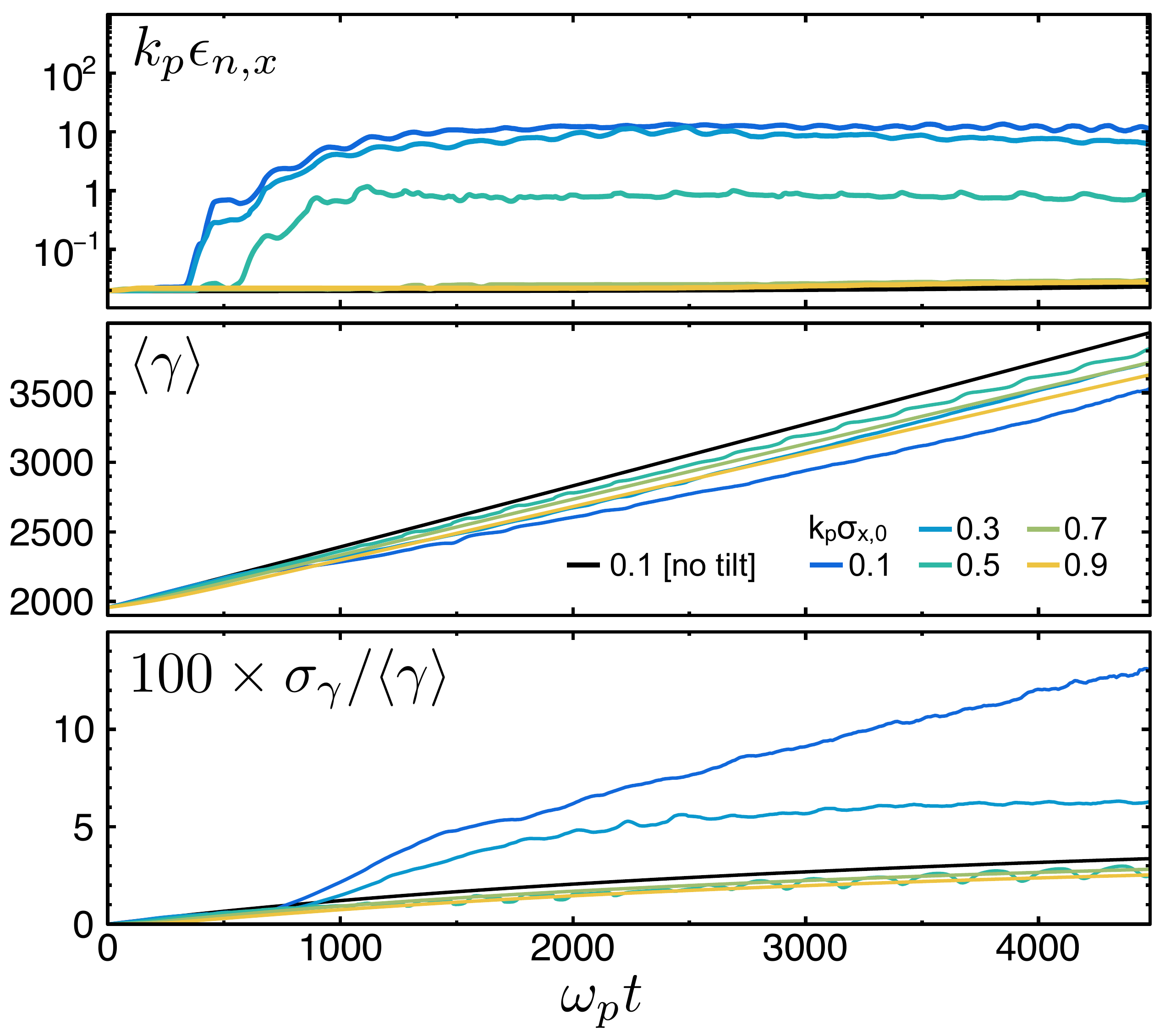}
  \caption{
    (Top) Average slice emittance, (middle) average energy and (bottom) relative energy spread of
    the witness beam as a function of the propagation time, for five PIC simulation cases
    with different initial driver transverse size and same initial emittance $k_p \epsilon_{x,0} = 0.02$.
    An ideal narrow case unaffected by hosing (with no initial tilt) is also included as a reference.
  }
\label{fig:FIG6} 
\end{figure*}

Fig.~\ref{fig:FIG5} shows the average slice emittance of the witness beam
as a function of the propagation time,
for five PIC simulation cases with different initial emittance and same initial transverse size
of the drive beam, which ranges from $k_p\sigma_{x,0} = 0.1$ to $k_p\sigma_{x,0} = 0.9$
(from top to bottom).
Fig.~\ref{fig:FIG6} shows the time evolution of the average sliced emittance (top panel),
the average energy (middle panel) and the relative energy spread (bottom panel) 
of the witness beam, for five cases with increasing initial drive beam size.
Only for the initially wide beam cases with $k_p\sigma_{x,0} = 0.7$ and $k_p\sigma_{x,0} = 0.9$,
the hosing instability could be rapidly suppressed and 
the witness beams could be efficiently accelerated with no emittance growth.
The acceleration performance in terms of the achieved energy gain in the witness beam
is $~10\%$ less for the case with $k_p\sigma_{x,0} = 0.7$
and $~15\%$ less for the case with $k_p\sigma_{x,0} = 0.9$,
with respect to a reference case with $k_p\sigma_{x,0}=0.1$ and no initial tilt (i.e. no hosing seed).
The relative energy spread of the witness beam after a propagation time $\omega_pt = 4500$
is $2.82\%$ for the case with $k_p\sigma_{x,0} = 0.7$,
$2.53\%$ for the case with $k_p\sigma_{x,0} = 0.9$
and $3.37\%$ for the reference case with $k_p\sigma_{x,0}=0.1$ and no initial tilt.

We have also included in the Supplemental Material online 
two PIC simulation movies showing the beam and electron densities
as a function of the propagation time, for two exemplary cases
with a narrow beam $k_p \sigma_{x,0} = 0.1$ (case $C_a$) and
a wide beam $k_p \sigma_{x,0} = 0.5$ (case $C_b$).

\end{document}